\documentclass[prd,twocolumn,notitlepage,showpacs,preprintnumbers,amsmath,amssymb,nofootinbib,APS,10pt,superscriptaddress]{revtex4-1}

\usepackage{dcolumn}
\usepackage{bm}
\usepackage{xcolor}
\usepackage[applemac]{inputenc}
\usepackage[spanish,english]{babel}
\usepackage{amsmath,amssymb,amsfonts,latexsym,cancel}
\usepackage{graphicx}
\usepackage{color}
\usepackage{soul}
\usepackage{ulem}
\usepackage{hyperref}
\usepackage{amsmath}
\usepackage{slashed}
\usepackage{float}

\allowdisplaybreaks

\newcommand{\be}{\begin{equation}}
\newcommand{\ee}{\end{equation}}
\newcommand{\bea}{\begin{eqnarray}}
\newcommand{\eea}{\end{eqnarray}}

\begin{document}

\title{{\bf $(\mathcal{F},\mathcal{G})$-summed form of the QED effective action}}

\author{Jose Navarro-Salas}\email{jnavarro@ific.uv.es}
\author{Silvia Pla}\email{silvia.pla@uv.es}

\affiliation{Departamento de Fisica Teorica and IFIC, Centro Mixto Universidad de Valencia-CSIC. Facultad de Fisica, Universidad de Valencia, Burjassot-46100, Valencia, Spain.}

\begin{abstract}

We conjecture that the proper-time series expansion of the one-loop effective Lagrangian of quantum electrodynamics 
can be  summed in all terms containing  the field-strength invariants $\mathcal{F} = \frac{1}{4} F_{\mu\nu}F^{\mu\nu} (x)$, $\mathcal{G}= \frac{1}{4} \tilde F_{\mu\nu}F^{\mu\nu}(x)$, including those also possessing derivatives of the electromagnetic field strength. This  partial resummation is exactly encapsulated in a factor with the same form as the Heisenberg-Euler Lagrangian density, except that now the electric and magnetic fields can depend arbitrarily on spacetime coordinates. 
We provide strong evidence for this conjecture, which is proved to sixth order in the proper time. Furthermore, and as a byproduct, we generate some solvable electromagnetic backgrounds. We  also discuss the implications for a generalization of the Schwinger formula for pair production induced by nonconstant electric fields.
 Finally, we briefly outline the extension of these results in the presence of gravity.

\end{abstract}


\date{\today}
\maketitle

\section{Introduction and main results}
The Heisenberg-Euler Lagrangian \cite{HE}  describes the nonlinearities of quantum electrodynamics (QED) when the fermionic degrees of freedom of matter  are integrated out and the strength of the electromagnetic  background is kept constant \cite{Dunne1,  Schwartz}.   At leading order, the quantum-corrected  term of the Heisenberg-Euler action  reads 
\be S^{(1)} = \frac{e^4 \hbar}{360 \pi^2 m_e^4 c^8} \int d^4x [(\vec E^2 - \vec B^2)^2 + 7 (\vec E  \vec B)^2] + \cdots \ee
where $m_e$ is the mass of the electron. 
The intrinsic nonlinearities of the quantum corrections have very important implications: light-by-light scattering, vacuum polarization, pair creation from vacuum, etc (see Ref. \cite{dunne3} and references therein).
A similar action can also be constructed for scalar QED \cite{Weisskopf}. Both theories, when reexpressed in the modern language of quantum field theory \cite{Schwinger},   can be regarded as  the 
archetypical models  of effective field theories (see also Ref. \cite{Vanyashin-Terentev} for a charged vector field theory).  
Furthermore, the phenomena of spontaneous pair creation by electric fields is linked to the imaginary part of the effective action, as first shown by Schwinger for a  constant electric field. Since the rate of particle production \cite{Schwinger} has an essential singularity in the electric charge, it is also regarded as the prototype of an intrinsically nonperturbative effect. It could be on the verge of being experimentally detected \cite{dunne2}.   Spontaneous particle creation is also fundamental in cosmology and black holes physics \cite{parker66, hawking, parker-toms,fulling, birrell-davies}. 


For arbitrary background configurations the form of the effective action is generically unknown, up to some specific exactly solvable cases \cite{Dunne1}. 
However, one can construct a generic asymptotic expansion for the one-loop effective action for scalar/spinor QED 
 based on the Schwinger proper-time expansion of the Feynman propagator \cite{Schwinger}. 
  It is closely connected to the  heat-kernel expansion and related techniques \cite{Vassilevich}.  More precisely,
the propagator can be expressed as (from now on we take $c=1=\hbar$)

\be G_F(x, x')= -i \int_0^\infty ds e^{-im^2s} \langle x, s| x', 0 \rangle \ , \ee 
with the kernel $ \langle x, s| x', 0 \rangle$ obeying an imaginary time diffusion equation 
with  appropriate boundary condition. The ``transition amplitude'' $ \langle x, s| x', 0 \rangle$ admits an asymptotic  expansion in powers of ``proper time'' $s$ which translates, at coincidence $x=x'$, into an expansion of the one-loop effective action $S^{(1)}$ 
\be S^{(1)} =  -\frac{i}{2} \int d^4x  \int_0^\infty \frac{ds}{s} e^{-im^2 s} \operatorname{tr} \langle x, s| x, 0 \rangle \ . \label{Skernel} \ee

The role of the classical electromagnetic background can be replaced by the gravitational field, which is naturally coupled to quantized  matter fields \cite{DeWittbook}. In general $\langle x, s| x, 0 \rangle \equiv \frac{i}{(4\pi is)^{2}} f(x,x;is)$ can be  expanded as $ \langle x, s| x, 0 \rangle = \frac{i}{(4\pi is)^{2}} \sum_{n=0}^\infty (is)^n a_n(x) $,  where  the DeWitt coefficients $a_n$ 
are local, covariant, and gauge-invariant quantities of mass dimension $2n$ \cite{DeWittbook, gilkey} (see also Refs. \cite{birrell-davies,fulling, parker-toms, Vassilevich}). 
In the purely gravitational  context, there is also a special soluble case, namely, the static Einstein universe, for which the Feynman propagator, and hence the effective action, can be evaluated exactly \cite{Bekenstein-Parker}. Bekenstein and Parker found that $ \langle x, s| x', 0 \rangle$ can be expressed as a Gaussian path integral, leading to   
$  \langle x, s| x, 0 \rangle= \frac{i}{(4\pi is)^{2}}e^{-iR(x)(\xi-\frac{1}{6})s}$, where $\xi$ is the coupling of the real scalar field with the Ricci scalar $R$. This is a nonperturbative result since it involves all powers of $s$ ($s \to 0$ represents the ultraviolet region, while $s\to \infty$ describes the infrared behavior). The remarkable point here is that the above solution captures the exact dependence on the Ricci scalar in the generic Schwinger-DeWitt, or heat-kernel, expansion. The general form of the kernel admits the following factorization:
\be  \langle x, s| x, 0 \rangle= \frac{i}{(4\pi is)^{2}}e^{-iR(x)(\xi-\frac{1}{6})s} \bar f(x,x;is)  \ , \ee
where the proper-time series of $\bar f= \sum_{n=0}^\infty (is)^n\bar a_n(x)$ contain no terms which vanish when $R(x)$ is replaced by zero \cite{parker-toms}.
This was first conjectured by Parker and Toms \cite{Parker-Toms85}, providing evidence to third order in the proper time 
(for a detailed proof see Ref. \cite{Jack-Parker} and for the counterpart in adiabatic regularization see Ref. \cite{FNP20}). Therefore, the quantum piece of the gravitational effective Lagrangian can be further expressed as 
 \be \mathcal{L}_{gravity}^{(1)} =  \frac{1}{2}  \int_0^\infty \frac{ds}{s}   \frac{e^{-im^2 s}}{(4\pi is)^{2}}e^{-iR(x)(\xi-\frac{1}{6})s} 
  \sum_{n=0}^\infty (is)^n\bar a_n(x) \label{Skernel2} \ \ \ \ee
This factorization of the nonperturbative term $e^{-iR(x)(\xi-\frac{1}{6})s}$ has major physical consequences to account for the effective dynamics of the Universe and the observed cosmological acceleration \cite{parker-raval,parker-ravalA, parker-vanzella, caldwell-komp-parker-vanzella, Melchiorri, Tommi, prl, JHEP}, and also for curvature dependence in the running of the gauge coupling constants \cite{CJP}.

In this Letter we point out a somewhat similar factorization for the electromagnetic interaction. We formulate the following conjecture: {\it the proper-time asymptotic expansion of the QED effective Lagrangian admits an exact resummation  in all terms involving the field-strength invariants $\mathcal{F}(x) = \frac{1}{4} F_{\mu\nu}F^{\mu\nu}(x) $, $\mathcal{G}(x)= \frac{1}{4} \tilde F_{\mu\nu}F^{\mu\nu}(x)$. The form of the factor involved in this partial resummation is just the Heisenberg-Euler Lagrangian for  QED, where the electric and magnetic fields depend  arbitrarily on spacetime coordinates.} 
We provide strong evidence of the validity of this conjecture, for both scalar and spinor QED. For simplicity we restrict ourselves to  four-dimensional Minkowski spacetime.

 Since our partial resummation involves two quantities, $\mathcal{F}(x)$ and $\mathcal{G}(x)$,  instead of the  single Ricci scalar $R(x)$,  the  factorization property described here  is more involved than the factorization of the exponential term found in gravity.  We find that (for computational purposes we find it very useful to use the notation of Ref. \cite{parker-toms}) 
 \bea    \label{Lscalar} \mathcal {L}_{scalar}^{(1)}&=& \int_0^\infty \frac{ds}{s} e^{-im^2 s}  \Bigg [\det \left( \frac{esF(x)}{\sinh (esF(x))}\right )\Bigg]^{1/2}   \\
&& \times \bar g(x;is)  \ , \nonumber \eea 
 \bea  \mathcal {L}_{spinor}^{(1)}&=&-\frac{1}{2} \int_0^\infty \frac{ds}{s} e^{-im^2 s}  \Bigg[\det \left( \frac{esF(x)}{\sinh (esF(x))}\right )\Bigg]^{1/2} \nonumber  \\ 
&&\times \operatorname{tr} [e^{-\frac{1}{2} es F_{\mu\nu}\sigma^{\mu\nu}}] \bar h(x;is)\ , \label{Lspinor}\eea with 
$F(x)\equiv F^{\mu}_{\, \, \nu}(x)$ and $\sigma^{\mu \nu}=\frac{1}{2}[\gamma^{\mu},\gamma^{\nu}]$. While the factors $\Big [\det \left( \frac{esF(x)}{\sinh (esF(x))}\right )\Big]^{1/2}$ and  $\operatorname{tr} [e^{-\frac{1}{2} es F_{\mu\nu}\sigma^{\mu\nu}}]$ are  functions depending only on  $\mathcal{F}(x)$ and $\mathcal{G}(x)$, the proper-time series for $\bar g(x;is)$ and $\bar h(x;is)$ contain no terms which vanish when  $\mathcal{F}(x)$ and $\mathcal{G}(x)$ 
(but not their derivatives) are replaced by zero. Therefore, all dependence on $\mathcal{F}(x), \mathcal{G}(x)$ has been encapsulated in the Heisenberg-Euler type factors of the above expansions. 
 Furthermore, our conjecture also has a direct extension to curved spacetime. Then the partial resummation involves the three basic scalars: $\mathcal{F}(x)$,  $\mathcal{G}(x)$, and $R(x)$.

Our computations rely on previous results for the Schwinger-DeWitt expansion of the effective action obtained within the so-called string-inspired world-line formalism \cite{coefs1, coefs2, coefs3,Schubert1} (see also Refs. \cite{string-inspired1, string-inspired2}). 
\\

\section{Scalar QED}
Let us consider a  quantized complex scalar field  $\phi$ coupled to an electromagnetic background field. 
The scalar field satisfies the equation
$
(D_{\mu}D^{\mu}+m^2)\phi=0
$, 
where $D_\mu=\partial_\mu - i e A_\mu$. The one-loop effective action of the scalar field $S^{(1)}_{scalar}$  admits an adiabatic expansion, of the form (\ref{Skernel}), 
consisting of an expansion in the number of external fields and the number of derivatives. (It is also related to the so-called large mass or inverse mass expansion.)  This expansion can be reexpressed, using the string-inspired method in the world-line formalism \cite{coefs1}, in the form 
\bea
S_{scalar}^{(1)}= -i \int d^4x \int_0^\infty\frac{ds}{s} e^{-ism^2} \,g(x;is) \label{Se2}
\eea
where ($O_0= 1$) 
\be
g(x;is)=\frac{i} {(4\pi is)^{2}}\sum_{n=0}^{\infty}\frac{(-is)^{n}}{n!}O_n(x).
\ee
$g(x;is)$ is directly related to the heat kernel, up to total derivatives 
\be g(x;is) = \langle x, s| x, 0 \rangle + total \ derivatives. \ee  
The advantage 
of this expansion is that it is possible to find a minimal basis for the coefficients $O_n(x)$ using the Bianchi identity, the antisymmetry of $F_{\mu \nu}$ and also integration by parts \cite{Muller}.
In this basis, the gauge-invariant coefficients of the expansion $O_n$ are written in the most compact form possible. 
The coefficients $O_n$ have been obtained up to 
twelfth adiabatic order ($n=6$) \cite{coefs1}. The expressions given in Ref. \cite{coefs1}  are also valid for non-Abelian gauge backgrounds and  also include a matrix-valued scalar potential $V(x)$. The first five coefficients of this expansion are given by 

{\small \bea
O_{1}&=&0, \,\,\,  O_{2}=- \tfrac{e^2}{6} F_{\kappa \lambda } F^{\kappa \lambda }, \,\,\,O_{3}=- \tfrac{e^2}{20} \partial_{\mu }F_{\kappa \lambda } \partial^{\mu }F^{\kappa \lambda },\\
O_{4}&=&\tfrac{e^4}{15} F_{\kappa }{}^{\mu } F^{\kappa \lambda } F_{\lambda }{}^{\nu } F_{\mu \nu } + \tfrac{e^4}{12} F_{\kappa \lambda } F^{\kappa \lambda } F_{\mu \nu } F^{\mu \nu } \\&-&  \tfrac{e^2}{70} \partial_{\nu }\partial_{\mu }F_{\kappa \lambda } \partial^{\nu }\partial^{\mu }F^{\kappa \lambda },\nonumber\\
O_5&=&\tfrac{2e^4}{7} F^{\kappa \lambda } F^{\mu \nu } \partial_{\lambda }F_{\nu \rho } \partial_{\mu }F_{\kappa }{}^{\rho } -  \tfrac{4e^4}{63} F_{\kappa }{}^{\mu } F^{\kappa \lambda } \partial_{\lambda }F^{\nu \rho } \partial_{\mu }F_{\nu \rho } \\
&&-  \tfrac{e^4}{9} F_{\kappa }{}^{\mu } F^{\kappa \lambda } F^{\nu \rho } \partial_{\mu }\partial_{\lambda }F_{\nu \rho } -  \tfrac{16e^4}{63} F^{\kappa \lambda } F^{\mu \nu } \partial_{\mu }F_{\kappa }{}^{\rho } \partial_{\nu }F_{\lambda \rho } \nonumber\\
&&+ \tfrac{5e^4}{18} F^{\kappa \lambda } F^{\mu \nu } \partial_{\rho }F_{\mu \nu } \partial^{\rho }F_{\kappa \lambda } + \tfrac{34e^4}{189} F^{\kappa \lambda } F^{\mu \nu } \partial_{\nu }F_{\lambda \rho } \partial^{\rho }F_{\kappa \mu } \nonumber\\
&&+ \tfrac{25e^4}{189} F^{\kappa \lambda } F^{\mu \nu } \partial_{\rho }F_{\lambda \nu } \partial^{\rho }F_{\kappa \mu } + \tfrac{4e^4}{21} F_{\kappa }{}^{\mu } F^{\kappa \lambda } \partial_{\rho }F_{\mu \nu } \partial^{\rho }F_{\lambda }{}^{\nu } \nonumber\\
&&+ \tfrac{e^4}{12} F_{\kappa \lambda } F^{\kappa \lambda } \partial_{\rho }F_{\mu \nu } \partial^{\rho }F^{\mu \nu } -  \tfrac{e^2}{252} \partial_{\rho }\partial_{\nu }\partial_{\mu }F_{\kappa \lambda } \partial^{\rho }\partial^{\nu }\partial^{\mu }F^{\kappa \lambda }\nonumber.
\eea}
 We remark again that the above proper-time expansion can  be regarded as an adiabatic expansion, with the  adiabatic assignment $1$ for $A_\mu$, $2$ for $F_{\mu\nu}$, $3$ for $\partial_\rho F_{\mu\nu}$, etc. The adiabatic order for $O_n$ is $2n$.  This  is equivalent to grouping terms with a fixed mass dimension. This expansion is different from a purely derivative expansion of the effective action, as given for instance in Ref. \cite{igor}. 

Our claim is that we can make a partial resummation  of the proper-time asymptotic expansion of the effective action, factorizing all terms containing the field-strength invariants $\mathcal{F}(x)$ and $\mathcal{G}(x)$, namely
\bea \label{U}g(x;is) &=& \Big[\det\bigg(\frac{es F(x)}{\sinh(e s F(x))}\bigg)\Big]^{1/2}  \bar g(x;is)\\
&\equiv& U(x;is) \,  \bar g(x;is), \nonumber \eea
where $\bar g(x;is)$ can be adiabatically expanded as 
\be
 \bar g(x;is)=\frac{i} {(4\pi is)^{2}}\sum_{n=0}^\infty \frac{ \bar O_n}{n!}(-i s )^n.
\label{gexp} \ee
 The first terms of the adiabatic/proper-time expansion of the resummed function are
 {\small 
 \be U(x;is)\sim 1+ U_2(x)(-i s)^2+ U_4(x)(-i s)^4+ U_6(x)(-i s)^6+\cdots , \label {det-expansion}
\ee}
where 
{\small \bea
U_2(x)&=&\tfrac{e^2}{12}\operatorname{Tr}(F^2),\\
U_4(x)&=&\tfrac{e^4}{288}\operatorname{Tr}(F^2)^2+\tfrac{e^4}{360}\operatorname{Tr}(F^4),\\
U_6(x)&=&\tfrac{e^6}{10368}\operatorname{Tr}(F^2)^3+\tfrac{e^6 }{4320}\operatorname{Tr}(F^2) \operatorname{Tr}(F^4)+\tfrac{e^6}{5670}\operatorname{Tr}(F^6).\nonumber \\
\eea}
Combining the expansions above we can immediately obtain the form of the $\bar O_n$ adiabatic coefficients. For the first terms we find 
{\small$\bar O_1=\bar O_2=0$}, {\small$\bar O_3=O_3$}, 
{\small \bea
\bar O_4&=&-  \tfrac{e^2}{70} \partial_{\nu }\partial_{\mu }F_{\kappa \lambda } \partial^{\nu }\partial^{\mu }F^{\kappa \lambda },\\
 \bar O_5&=&\tfrac{2e^4}{7} F^{\kappa \lambda } F^{\mu \nu } \partial_{\lambda }F_{\nu \rho } \partial_{\mu }F_{\kappa }{}^{\rho } -  \tfrac{4e^4}{63} F_{\kappa }{}^{\mu } F^{\kappa \lambda } \partial_{\lambda }F^{\nu \rho } \partial_{\mu }F_{\nu \rho } \\
&&-  \tfrac{e^4}{9} F_{\kappa }{}^{\mu } F^{\kappa \lambda } F^{\nu \rho } \partial_{\mu }\partial_{\lambda }F_{\nu \rho } -  \tfrac{16e^4}{63} F^{\kappa \lambda } F^{\mu \nu } \partial_{\mu }F_{\kappa }{}^{\rho } \partial_{\nu }F_{\lambda \rho } \nonumber\\
&&+ \tfrac{5e^4}{18} F^{\kappa \lambda } F^{\mu \nu } \partial_{\rho }F_{\mu \nu } \partial^{\rho }F_{\kappa \lambda } + \tfrac{34e^4}{189} F^{\kappa \lambda } F^{\mu \nu } \partial_{\nu }F_{\lambda \rho } \partial^{\rho }F_{\kappa \mu } \nonumber\\
&&+ \tfrac{25e^4}{189} F^{\kappa \lambda } F^{\mu \nu } \partial_{\rho }F_{\lambda \nu } \partial^{\rho }F_{\kappa \mu } + \tfrac{4e^4}{21} F_{\kappa }{}^{\mu } F^{\kappa \lambda } \partial_{\rho }F_{\mu \nu } \partial^{\rho }F_{\lambda }{}^{\nu } \nonumber\\
&&-  \tfrac{e^2}{252} \partial_{\rho }\partial_{\nu }\partial_{\mu }F_{\kappa \lambda } \partial^{\rho }\partial^{\nu }\partial^{\mu }F^{\kappa \lambda }\nonumber .
\eea}
{\it Note that  the $\bar O_n$ coefficients do not contain any terms going as $\mathcal{F}(x)$ and $\mathcal{G}(x)$,} or, equivalently, they do not contain any terms proportional to {\small $\operatorname{Tr}(F^2)$, $\operatorname{Tr}(F^4)$, $\operatorname{Tr}(F^6) \cdots .$}
We have verified our conjecture  to sixth order in proper time. 
The expression for  $O_6$ contains $41$ terms and it is, as far as we know,  the highest  available coefficient in the literature. 

 If we include an additional scalar field background $\Phi(x)$ the quantized charged scalar satisfies $(D_\mu D^{\mu}+m^2+\Phi(x))\phi=0$.  We have verified that the factorization of the Heisenberg-Euler type factor also occurs. Furthermore, we find that in this case we can also factorize an exponential term, that is, 
\be g(x;is) = \Bigg[\det\bigg(\frac{es F(x)}{\sinh(e s F(x))}\bigg)\Bigg]^{1/2} e^{-is\Phi(x)} \bar g(x;is). \label{scalar-F-factor}\ee
The exponential factorization was also found in Refs. \cite{Jack-Parker, coefs3} without the electromagnetic background. For the case given in Eq. \eqref{scalar-F-factor} $O_6$ has 97 terms while $\bar O_6$ has 62. 

 We note finally that a purely derivative expansion \cite{igor} does not satisfy  the factorization property obtained within the proper-time expansion.\\

\section{Spinor QED}
Let us consider now a  quantized charged spin-$\frac{1}{2}$ field $\psi$. The second-order equation for the spinor field is  $(D_\mu D^\mu+m^2-\frac{i}{2}eF_{\mu \nu}\sigma^{\mu \nu})\psi=0$. 
As in the scalar case, the induced one-loop effective action  admits the following adiabatic expansion:
 \bea  \mathcal {S}_{spinor}^{(1)}&=&\frac{i}{2} \int d^4x \int_0^\infty \frac{ds}{s} e^{-im^2 s} h(x;is) \ , \eea
with 
 \bea
h(x;is)&=&\frac{i} {(4\pi is)^{2}}\operatorname{tr}\sum_{n=0}^{\infty}\frac{(-is)^{n}}{n!}O_n(x)\\ &\equiv& \frac{i} {(4\pi is)^{2}}\sum_{n=0}^{\infty}\frac{(-is)^{n}}{n!}o_n(x) ,
 \nonumber \eea
and where we have defined $o_n=\operatorname{tr}O_n$. Note that $O_0= I_{4\times 4}$ and $o_0=4$. 
Again, we will use the expansion given in Ref. \cite{coefs1} (with the appropriate redefinitions). 
 The leading terms of this adiabatic expansion are
{\small \bea
o_{1}&=&0, \,\,\,  o_{2}=\tfrac{4e^2}{3} F_{\kappa \lambda } F^{\kappa \lambda }, \,\,\,o_{3}=\tfrac{4e^2}{5} \partial_{\mu }F_{\kappa \lambda } \partial^{\mu }F^{\kappa \lambda },\\
o_{4}&=&- \tfrac{56e^4}{15} F_{\kappa }{}^{\mu } F^{\kappa \lambda } F_{\lambda }{}^{\nu } F_{\mu \nu } + \tfrac{4e^4}{3} F_{\kappa \lambda } F^{\kappa \lambda } F_{\mu \nu } F^{\mu \nu }\\&& + \tfrac{12e^{2}}{35} \partial_{\nu }\partial_{\mu }F_{\kappa \lambda } \partial^{\nu }\partial^{\mu }F^{\kappa \lambda },\nonumber \\
o_5&=&\tfrac{8e^4}{7} F^{\kappa \lambda } F^{\mu \nu } \partial_{\lambda }F_{\nu \rho } \partial_{\mu }F_{\kappa }{}^{\rho } -  \tfrac{16e^4}{63} F_{\kappa }{}^{\mu } F^{\kappa \lambda } \partial_{\lambda }F^{\nu \rho } \partial_{\mu }F_{\nu \rho } \\
&&+ \tfrac{8e^4}{9} F_{\kappa }{}^{\mu } F^{\kappa \lambda } F^{\nu \rho } \partial_{\mu }\partial_{\lambda }F_{\nu \rho } -  \tfrac{232e^4}{63} F^{\kappa \lambda } F^{\mu \nu } \partial_{\mu }F_{\kappa }{}^{\rho } \partial_{\nu }F_{\lambda \rho } \nonumber \\
&&+ \tfrac{40e^4}{9} F^{\kappa \lambda } F^{\mu \nu } \partial_{\rho }F_{\mu \nu } \partial^{\rho }F_{\kappa \lambda } + \tfrac{136e^4}{189} F^{\kappa \lambda } F^{\mu \nu } \partial_{\nu }F_{\lambda \rho } \partial^{\rho }F_{\kappa \mu } \nonumber \\
&&-  \tfrac{656e^4}{189} F^{\kappa \lambda } F^{\mu \nu } \partial_{\rho }F_{\lambda \nu } \partial^{\rho }F_{\kappa \mu } -  \tfrac{320e^4}{21} F_{\kappa }{}^{\mu } F^{\kappa \lambda } \partial_{\rho }F_{\mu \nu } \partial^{\rho }F_{\lambda }{}^{\nu } \nonumber \\
&&+ \tfrac{8e^4}{3} F_{\kappa \lambda } F^{\kappa \lambda } \partial_{\rho }F_{\mu \nu } \partial^{\rho }F^{\mu \nu } + \tfrac{8e^2}{63} \partial_{\rho }\partial_{\nu }\partial_{\mu }F_{\kappa \lambda } \partial^{\rho }\partial^{\nu }\partial^{\mu }F^{\kappa \lambda }.\nonumber 
\eea}
Here, the resummed expansion is given by
\bea \label{W}
 h(x;is) &=& \Big[\det\bigg(\frac{es F}{\sinh(e s F)}\bigg)\Big]^{1/2} \operatorname{tr}[e^{-e s \frac{1}{2}F_{\mu \nu} \sigma^{\mu \nu}}] \,\, \bar h(x;is) \ \ \nonumber   \\
 &\equiv& W(x;is)\, \bar h(x;is) \ ,  \label{f-factor}
 \eea
where
\bea
\bar h(x;is)= \frac{i} {(4\pi is)^{2}}\sum_{n=0}^{\infty}\frac{(-is)^{n}}{n!} \bar o_n(x) .
\label{hexp}\eea
In this case, the adiabatic expansion of the resummed function is given by
{\small 
 \be W(x;is)\sim
\operatorname{tr}{I}+ W_2(x)(-i s)^2+ W_4(x)(-i s)^4+ W_6(x)(-i s)^6+\cdots ,
\ee}
where 
{\small \bea
W_2(x)&=& -\tfrac{2e^2}{3}\operatorname{Tr}(F^2),\\ 
 W_4(x)&=&\tfrac{e^4}{18}\operatorname{Tr}(F^2)^2-\tfrac{ 7 e^4}{45}\operatorname{Tr}(F^4), \\
 W_6(x)&=& -\tfrac{e^6}{324}\operatorname{Tr}(F^2)^3+\tfrac{7e^6 }{270}\operatorname{Tr}(F^2) \operatorname{Tr}(F^4)-\tfrac{124e^6}{2835}\operatorname{Tr}(F^6). \nonumber \\
\eea}
Then, combining all previous expansions, we find the following coefficients for the resummed expansion,  $\bar o_0=1$, $\bar o_1=\bar o_2=0$, $4\,\bar o_3=o_3$, 
{\small \bea \label{fbarCoeff}
4 \,\bar o_{4}&=&\tfrac{12e^2}{35} \partial_{\nu }\partial_{\mu }F_{\kappa \lambda } \partial^{\nu }\partial^{\mu }F^{\kappa \lambda },\\
4\, \bar o_5&=&\tfrac{8e^4}{7} F^{\kappa \lambda } F^{\mu \nu } \partial_{\lambda }F_{\nu \rho } \partial_{\mu }F_{\kappa }{}^{\rho } -  \tfrac{16e^4}{63} F_{\kappa }{}^{\mu } F^{\kappa \lambda } \partial_{\lambda }F^{\nu \rho } \partial_{\mu }F_{\nu \rho } \\
&&+ \tfrac{8e^4}{9} F_{\kappa }{}^{\mu } F^{\kappa \lambda } F^{\nu \rho } \partial_{\mu }\partial_{\lambda }F_{\nu \rho } -  \tfrac{232e^4}{63} F^{\kappa \lambda } F^{\mu \nu } \partial_{\mu }F_{\kappa }{}^{\rho } \partial_{\nu }F_{\lambda \rho } \nonumber \\
&&+ \tfrac{40e^4}{9} F^{\kappa \lambda } F^{\mu \nu } \partial_{\rho }F_{\mu \nu } \partial^{\rho }F_{\kappa \lambda } + \tfrac{136e^4}{189} F^{\kappa \lambda } F^{\mu \nu } \partial_{\nu }F_{\lambda \rho } \partial^{\rho }F_{\kappa \mu } \nonumber \\
&&-  \tfrac{656e^4}{189} F^{\kappa \lambda } F^{\mu \nu } \partial_{\rho }F_{\lambda \nu } \partial^{\rho }F_{\kappa \mu } -  \tfrac{320e^4}{21} F_{\kappa }{}^{\mu } F^{\kappa \lambda } \partial_{\rho }F_{\mu \nu } \partial^{\rho }F_{\lambda }{}^{\nu } \nonumber \\
&& + \tfrac{8e^2}{63} \partial_{\rho }\partial_{\nu }\partial_{\mu }F_{\kappa \lambda } \partial^{\rho }\partial^{\nu }\partial^{\mu }F^{\kappa \lambda } \nonumber.
\eea}

We note here that we have proposed a factorization of the one-loop effective action given by Eq. \eqref{f-factor},  but this does not preclude the possibility of finding a somewhat similar factorization for the heat kernel itself, as it happens in the gravitational case. This  deserves further analysis.   Furthermore, the proposed form of the effective action could also be of interest to analyze the backreaction issue in the pair creation process (for a recent analysis in $1+1$ dimensions see Ref. \cite{Validity}). \\

\section{Exactly solvable electromagnetic backgrounds}
An involved test of the factorization property is provided by the exact one-loop effective action obtained for an electric field which points in the $\hat z$ direction and depends arbitrarily on the light-cone coordinate $x^+\equiv (t + z)$, as described for spinor QED in Refs. \cite{Woodard1, Woodard2}. In this case it is convenient to evaluate the factors $\bar g(x;is)$ and $\bar h(x;is)$ in Eqs. (\ref{Lscalar}) and (\ref{Lspinor}). We found that all coefficients of the expansions \eqref{gexp} and \eqref{hexp}  are zero, $\bar O_{n>0} =0=\bar o_{n>0} $. We checked this up to and including the sixth order in the proper time expansion. Then we get $\bar g(x;is)=\bar h(x;is)=i/(4\pi i s)^2$. 
This implies that, the (unrenormalized) quantum effective Lagrangians become
\bea     \mathcal {L}_{scalar}^{(1)}= \frac{-1}{16\pi^2} \int_0^\infty \frac{ds}{s^3} e^{-im^2 s}\frac{esE(x^+)}{\sinh esE(x^+)}   
 \label{Lscalar2} \eea 
 \bea  \mathcal {L}_{spinor}^{(1)}=\frac{1}{8\pi^2} \int_0^\infty \frac{ds}{s^3} e^{-im^2 s} \frac{esE(x^+)\cosh esE(x^+)}{\sinh esE(x^+)} 
\ . \  \label{Lspinor2}\eea 
Our result (\ref{Lspinor2}) agrees with that obtained in Ref. \cite{Woodard1}. Renormalization can be simply done by subtracting off all terms up to adiabatic order 4 (or, equivalently, up to second order in the proper-time expansion).

Our factorization conjecture 
can be used to easily predict the one-loop effective actions for some new families of  configurations for which $\bar O_{n>0}= \bar o_{n>0}=0$. 
In fact, we find this cancellation happens for electric and magnetic fields pointing in the $\hat z$ direction that depend arbitrarily on the light-cone coordinate $x^+=(t+ z)$. Then, the unrenormalized effective Lagrangians are given by 

\be     \mathcal {L}_{scalar}^{(1)}= \frac{-1}{16\pi^2} \int_0^\infty \frac{ds}{s^3} e^{-im^2 s}\frac{e^2s^2E(x^+)B(x^+)}{\sinh esE(x^+) \sin esB(x^+)}   
 \label{Lscalar3} \ee 
 \be  \mathcal {L}_{spinor}^{(1)}=\frac{1}{8\pi^2} \int_0^\infty \frac{ds}{s^3} e^{-im^2 s} \frac{e^2s^2E(x^+)B(x^+)}{\tanh esE(x^+)\tan{esB(x^+)}} 
\ . \  \label{Lspinor3}\ee 

This general result includes the case of a purely electric field explained above and also the case of a pure magnetic field of the form $\vec B=B(x^+)\hat z$. 
 The formulas above also encapsulate the case in which one of the fields is constant,  and  are consistent with the results found in Ref. \cite{Ilderton}. The same results \eqref{Lscalar3} and \eqref{Lspinor3} appear if we consider the coordinate dependence on $x^-=(t-z)$.


Another interesting example is the single plane wave field. As explicitly proved in Ref. \cite{Schwinger} the spinor one-loop effective action  vanishes. This result can be easily understood  from the factorization property in Eqs. (\ref{Lscalar}) and (\ref{Lspinor}). For a plane wave, the two Lorentz invariants $\mathcal{F}(x)$ and $\mathcal{G}(x)$ vanish, and hence  the corresponding effective actions. \\ 

\section{Generalization of Schwinger's formula for the pair production rate}
As a byproduct of our results we can estimate the imaginary part of the one-loop effective action 
according to Eq. (\ref{Lspinor}). Following standard arguments, it can  be obtained by rotating the integration contour from along the positive real axis to the negative imaginary axis ($i s \to \tau$).  The factorization found in this paper suggests that the  poles of the imaginary part of the one-loop effective action are located at the same points as in the constant electric field case $\tau_n= n\pi/e|\vec E(x)|$. 
This leads to the (proper-time) expansion for the pair creation rate
\bea \label{Schw-fermi} &&\operatorname{Im} S_{spinor}^{(1)}=\operatorname{Im} \int \frac{d^4x}{8\pi^2}\int_0^\infty \frac{ds}{s^3} e^{-im^2 s} \frac{esE(x)\cosh esE(x)}{\sinh esE(x)}\nonumber\\ &&\times \Big( 1 + \sum_{k=3}^{\infty}\frac{(-is)^{k}}{k!} \bar o_k(x)\Big) \\ 
&& = - 2 \pi i \int d^4x \sum_{n=1}^{\infty} \operatorname{Res}[\frac{e^{-m^2\tau}}{\tau} \frac{e\tau E(x)\cos e\tau E(x)}{\sin e\tau E(x)}  \bar h(x;\tau),\tau_n] \nonumber
\eea
where $\bar h (x;\tau)$ is given in Eq. \eqref{hexp} and $\bar o_k$ are the coefficients given in Eq. \eqref{fbarCoeff} for the case $F_{0i}=E_i(x)$ and $F_{ij}=0$. For instance, $\bar o_3=\tfrac{2}{5} ((\partial_jE^i(x))^2-(\partial_0E^i(x))^2)$.
The leading factor of the integrand reproduces Schwinger's result for the rate of electron-positron pair production, except that now the electric field can (slowly) vary on $x$. If one truncates the series expansion in Eq. \eqref{Schw-fermi} one gets perturbative weak-field corrections. To get significant (nonperturbative) corrections one has to sum all terms in the proper-time expansion containing all derivatives for a given number of fields, or sum over all power of fields with a given number of derivatives. The latter possibility has been given  for first derivative corrections via the derivative expansion worked out in Ref. \cite{igor}. 


 
 \section{Adding gravity}
Finally, we would like to stress that the above results appear to be robust against the gravitational interaction.\footnote{For $\nabla_\mu F^{\rho\sigma} =0$ the factorization of the Euler-Heisenberg action was given in Refs. \cite{Avramidi-Fucci1, Avramidi-Fucci2}.} In a curved spacetime  $\bar g(x,is)$ and $\bar h(x,is)$ can be further factorized as 
 $\bar g(x,is)= e^{-isR(x)(\xi-\frac{1}{6})} \tilde g(x, is)$, $\bar h(x,is)= e^{-\frac{is}{12}R(x)} \tilde h(x, is)$, where the expansion of $\tilde g(x, is)$ does not contain any term going as $\mathcal{F}(x)$, $\mathcal{G}(x)$, or $R(x)$. 
 This also has implications for the Schwinger pair creation formula since the exponential term in Eq. (\ref{Schw-fermi}) generalizes to 
 $\exp {[- is (m^2 + \frac{1}{12}R(x))]}$. 
 Furthermore, one of the consequences  of the exponential factorization of the scalar curvature in $\bar h(x, is)$ 
 is the emergence of a logarithmic correction of the form $\mathcal{F}(x) \log (1+ \frac{R(x)}{12m^2})$  to the effective action of QED in curved spacetime. The leading terms agree with the results in Refs. \cite{Drummod-Hathrell80, Schubert09}.  Note that we also have higher-order corrections proportional to $\frac{\tilde o_n}{(m^2 + R(x)/12)^{(n-2)}}$, with $n \ge 3$. We also expect (nonperturbative)  implications in the effective field theory studies of QED in curved spacetime, such as the issue concerning  the speed of light in the strong gravity regime \cite{CdR}.\\


 \section{Conclusions and final comments}
Inspired by the exact $R$-summed property of the effective action in gravity we have proposed a somewhat similar factorization in the proper-time form of the QED effective action in terms of the two field-strength invariants $\mathcal{F}$ and $\mathcal{G}$. 
The functional dependence of the conjectured resummation factor has the same functional form as that obtained by Heisenberg and Euler for constant electromagnetic fields. This is displayed in Eqs. \eqref{Lscalar} and \eqref{Lspinor}. We have checked this conjecture to sixth order in the proper-time expansion, and for both scalar and spinor electrodynamics. As a straightforward application of this proposal we have easily identified 
 families of electromagnetic backgrounds for which the effective action can be exactly solved. A generalization of the Schwinger result for pair production was also discussed. Furthermore, these results appear to be robust in the presence of a gravitational background. 

 We expect that a similar factorization of the effective Lagrangian density could also be found for quantized spin-1 fields and also for non-Abelian gauge backgrounds.  We leave this, and also the proof of the conjecture (along the lines of Ref. \cite{Jack-Parker}), for future studies. \\

\section*{Acknowledgments}
We thank P. R. Anderson, P. Beltr\'an-Palau, A. Ferreiro, G. Olmo and I. Shovkovy for useful comments and discussions. Most of the computations in this paper have been done with the help of the xAct package of the {\it Mathematica software}. 
The traces of products of gamma matrices  
 have been computed with the FeynCalc package of {\it Mathematica}  (references for the programs listed above can be found in Ref. \cite{programs}).  This work was supported in part by Spanish Ministerio de  Economia,  Industria  y  Competitividad  Grants  No. FIS2017-84440-C2-1-P (MINECO/FEDER, EU), No.  FIS2017-91161-EXP and the project PROMETEO/2020/079 (Generalitat Valenciana). S. P.  is supported by a Ph.D. fellowship, Grant No. FPU16/05287.


\begin{thebibliography}{99}

\bibitem{HE}
  W.~Heisenberg and H.~Euler,
 Z. Phys. {\bf 98}, 714 (1936). English translation: arXiv:physics/605038. 

\bibitem{Dunne1} G. D. Dunne, in {\it From Fields to Strings: Circumnavigating Theoretical Physics}, edited by M. Shifman, A. Vainshtein and J. Wheater (World Scientific, 2005), pp. 445-522; arXiv:hep-th/0406216.

\bibitem{Schwartz} M. D. Schwartz, {\it Quantum Field Theory and the Standard Model} (Cambridge University Press, Cambridge, England, 2014).

\bibitem{dunne3} G. V. Dunne,  Int. J. Mod. Phys. A {\bf 27}, 1260004 (2012);  F. Karbstein,  {\it Particles} {\bf 3}, 39 (2020).



\bibitem{Weisskopf} V. Weisskopf, The electrodynamics of the vacuum based on the quantum theory of the electron,  Kong. Dans. Vid. Selsk. Math-fys. Medd. {\bf 14N6} (1936).

\bibitem{Schwinger}
J. Schwinger, 
 Phys. Rev. {\bf 82}, 664 (1951).

 \bibitem{Vanyashin-Terentev} V.S. Vanyashin and M. V. Terent'ev,  Sov. Phys. JETP {\bf 21}, 375 (1965).


\bibitem{dunne2} G. V. Dunne,  Eur. Phys. J. D {\bf 55}, 327 (2009).







\bibitem{parker66} L. Parker,  Phys. Rev. Lett. {\bf 21}, 562 (1968);  Phys. Rev. {\bf 183}, 1057 (1969).

\bibitem{hawking} S. W. Hawking,  Commun. Math. Phys. {\bf 43}, 199 (1975).







\bibitem{birrell-davies} N. D. Birrell  and P.C.W. Davies, {\it Quantum Fields in Curved Space} (Cambridge University Press, Cambridge, England, 1982).


\bibitem{fulling} S. A. Fulling, {\it Aspects of Quantum Field Theory in Curved Space-Time} (Cambridge University Press, Cambridge, England, 1989).

\bibitem{parker-toms}L.~Parker and D.~J.~Toms, {\it Quantum Field Theory in Curved Spacetime: Quantized Fields
and Gravity} (Cambridge University Press, Cambridge, England, 2009).

\bibitem{Vassilevich}  D. V. Vassilevich,  Phys. Rep. {\bf 388}, 279 (2003). 





\bibitem{DeWittbook} B. S. DeWitt, {\it Dynamical Theory of Groups and Fields} (Gordon and Breach, New York, 1965). 

\bibitem{gilkey} P. B. Gilkey,  J. Differ. Geom.  {\bf 10}, 601 (1975).

\bibitem{Bekenstein-Parker} J.D. Bekenstein and L. Parker,  Phys. Rev. D {\bf 23}, 2850 (1981).




\bibitem{Parker-Toms85} L.~Parker and D.~J.~Toms,  Phys. Rev. D {\bf 31}, 953 (1985).

\bibitem{Jack-Parker} I. Jack and L. Parker,  Phys. Rev. D {\bf 31}, 2439 (1985).

\bibitem{FNP20} A. Ferreiro, J. Navarro-Salas and S. Pla,  Phys. Rev. D {\bf 101} 105011 (2020).

\bibitem{parker-raval} L. Parker and A. Raval,  Phys. Rev. D {\bf 60}, 063512 (1999). 
\bibitem{parker-ravalA} L. Parker and A. Raval,   Phys. Rev. D {\bf 62}, 083503 (2000).
\bibitem{parker-vanzella} L. Parker and D. A. T. Vanzella,  Phys. Rev. D {\bf 69}, 104009 (2004).
\bibitem{caldwell-komp-parker-vanzella} R.R. Caldwell, W. Komp, L. Parker, and D.A.T. Wanzella,  Phys. Rev. D {\bf 73}, 023513 (2006).

\bibitem{Melchiorri} 	 E. Di Valentino, E. V. Linder, and A. Melchiorri,  Phys. Rev. D {\bf 97},   043528 (2018).

\bibitem{Tommi} T. Markkanen, S. Nurmi, A. Rajantie, and  S. Stopyr,   J. High Energy Phys. 06 (2018)  040.
\bibitem{prl} 	E. V. Castro, A. Flachi, P. Ribeiro, and V. Vitagliano,  Phys. Rev. Lett.  {\bf 121}, 221601 (2018).
\bibitem{JHEP} 	O. Czerwinska, Z. Lalak, and L. Nakonieczny, J. High Energy Phys. 11 (2015) 207.

 \bibitem{CJP} E. Calzetta, I. Jack, and L. Parker,  Phys. Rev. Lett. {\bf 55}, 1241 (1985).




\bibitem{coefs1} D. Fliegner, P. Haberl, M.G. Schmidt and C. Schubert,  Ann. Phys. (N.Y.) {\bf264}, 51 (1998).

\bibitem{coefs2} D. Fliegner, M.G. Schmidt and C. Schubert,
Nucl. Instrum. Methods Phys. Res., Sect. A {\bf 389}, 374 (1997).



\bibitem{coefs3} D. Fliegner, M. G. Schmidt and C. Schubert,  Z. Phys. C {\bf 64}, 111 (1994).

\bibitem{Schubert1} M. Reuter, M. G. Schmidt, C. Schubert, Ann. Phys. (N.Y.) {\bf259}, 313 (1997).

\bibitem{string-inspired1} M. G. Schmidt and C. Schubert,  Phys. Lett. B {\bf 318}, 438 (1993).

\bibitem{string-inspired2} Z. Bern and D. A. Kosower,  Phys. Rev. Lett. {\bf 66}, 1669 (1991).



\bibitem{Muller} U. M\"uller, {\it New Computing Techniques in Physics Research IV} (World Scientific, Singapore, 1996), p. 193; DESY-96-154, arXiv:hep-th/9701124.


\bibitem{igor} V. P. Gusynin and I. A. Shovkovy,  J. Math. Phys. (N.Y.) {\bf 40}, 5406 (1999). 




 \bibitem{Validity} S. Pla, I. M. Newsome, R. S. Link, P. R. Anderson and J. Navarro-Salas, Validity of the semiclassical approximation for pair production due to an electric field in 1+1 dimensions, arXiv: 2010.09811  [Phys. Rev. D (to be published)].
 
\bibitem{Woodard1}H. M. Fried and R. P. Woodard,  Phys. Lett. B {\bf 524}, 233 (2002).
\bibitem{Woodard2}T. N. Tomaras, N. C. Tsamis, and R. P. Woodard, Phys. Rev. D {\bf 62}, 125005 (2000).


\bibitem{Ilderton} A. Ilderton,  J. High Energy Phys. 09 (2014) 166. 



\bibitem{Avramidi-Fucci1} I.G. Avramidi and G. Fucci,  Commun. Math. Phys. {\bf 291}, 543 (2009).

\bibitem{Avramidi-Fucci2} G. Fucci and I.G. Avramidi,  On the Gravitationally Induced Schwinger Mechanism,   in {\it 9th Conference on Quantum Field Theory under the Influence of External Conditions (QFEXT 09)}, pp. 485-491 [arXiv:0911.1099].

\bibitem{Drummod-Hathrell80} I.T. Drummond and S.J. Hathrell,  Phys. Rev. D {\bf 22}, 343 (1980).

\bibitem{Schubert09} F. Bastianelli, J.M. Davila, and C. Schubert, J. High Energy Phys.
 03 (2009) 086. 
\bibitem{CdR} C. de Rham and A.J. Tolley,  Phys. Rev. D {\bf 102}, 084048 (2020).  

\bibitem{programs} J. M. Mart\'in-Garc\'ia, xAct: Efficient tensor computer algebra for the Wolfram language, \url{ https://www.xAct.es}; 
R. Mertig, M. B\"ohm, and A. Denner, Comput. Phys. Commun., {\bf 64}, 345, (1991).










\end{thebibliography}
\end{document}